\documentclass{article}
\usepackage{spconf,amsmath,graphicx,multicol}
\usepackage{amsmath,graphicx,multicol}
\usepackage{amsfonts}
\usepackage{color}
\usepackage{bbm}
\usepackage{cite}
\usepackage{amssymb}
\usepackage{algorithm}
\usepackage{algorithmic}
\usepackage{hyperref}
\usepackage{tabularx}
\usepackage[utf8]{inputenc}
\usepackage[english]{babel}

\usepackage{amsthm}
\usepackage[keeplastbox]{flushend}
\usepackage{subcaption}
\usepackage[font={footnotesize}]{caption}
\usepackage{url}
\usepackage{mathtools}
\usepackage{comment}
\usepackage{float}

\newtheorem{proposition}{Proposition}[]
\newtheorem{theorem}{Theorem}[]

\DeclareMathOperator*{\argmax}{arg\,max}

\DeclareMathOperator{\E}{\mathbb{E}}

\def\n{{\mathbf n}}
\def\y{{\mathbf y}}

\def\C{{\mathbf C}}

\def\P{{\mathbf P}}
\def\I{{\mathbf I}}

\def\r{{\mathbf r}}

\def\x{{\mathbf x}}

\def\Y{{\mathbf Y}}
\def\A{{\mathbf A}}
\def\Q{{\mathbf Q}}

\def\U{{\mathbf U}}
\def\V{{\mathbf V}}
\def\R{{\mathbbm{R}}}

\def\I{{\mathbf I}}

\def\u{{\mathbf u}}

\def\S{{\cal S}}

\def\K{{\cal K}}

\newcommand{\ts}{\textsuperscript}

\let\oldref\ref
\renewcommand{\ref}[1]{(\oldref{#1})}
\newcommand{\RNum}[1]{\uppercase\expandafter{\romannumeral #1\relax}}
\makeatletter
\renewcommand{\fnum@figure}{Fig.~\thefigure}
\makeatother

\title{A Novel Scheme for Support Identification and Iterative Sampling of Bandlimited Graph Signals}
\name{Abolfazl Hashemi$^{\dag}$, Rasoul Shafipour$^{\ddagger}$, Haris Vikalo$^{\dag}$, and Gonzalo Mateos$^{\ddagger}$\thanks{Work in this paper was supported in part by the NSF award CCF-1750428.}}
\address{$^{\dag}$Dept. of Electrical and Computer Engineering,  University of Texas at Austin, Austin, TX, USA\\
$^{\ddagger}$Dept. of Electrical and Computer Engineering, University of Rochester, Rochester, NY 14627, USA}
\begin{document}
\ninept
\maketitle
\begin{abstract}
We study the problem of sampling and reconstruction of bandlimited graph signals where the objective is to select a node subset of prescribed cardinality that ensures interpolation of the original signal with the lowest reconstruction error. We propose an efficient iterative selection sampling approach and show that in the noiseless case the original signal is exactly recovered from the set of selected nodes. In the case of noisy measurements, a bound on the reconstruction error of the proposed algorithm is established. We further address the support identification of the bandlimited signal with unknown support and show that under a pragmatic sufficient condition, the proposed framework requires minimal number of samples to perfectly identify the support. The efficacy of the proposed methods are illustrated through numerical simulations on synthetic and real-world graphs.
\end{abstract}
\vspace{-0.2cm}
\section{Introduction}\label{sec:intro}
Network data naturally supported on the vertices of a graph are becoming increasingly ubiquitous, with examples ranging from measurements of neural activities at different regions of the brain \cite{huang2016graph}, to vehicle trajectories over road networks \cite{deri_nyc_taxi}.
 Under the assumption that properties of the network process relate to the underlying graph, the goal of graph signal processing (GSP) is to broaden the scope of traditional signal processing tasks and develop algorithms that fruitfully exploit this relational structure \cite{shuman2013,sandryhaila2013}.

A cornerstone generalization which has recently drawn considerable attention pertains to sampling 
and reconstruction of graph signals \cite{shomorony2014sampling,tsitsvero2016signals,anis2016efficient,chepuri2016subsampling,marques2016sampling,gama2016rethinking,chamon2017greedy,hashemi2018sampling,chen2015discrete}. The task of finding an exact sampling set to perform reconstruction with minimal information loss is known to be NP-hard and conditions for exact reconstruction of graph signals from noiseless samples were put forth in~\cite{shomorony2014sampling,tsitsvero2016signals,anis2016efficient,chen2015discrete}. Existing approaches for sampling and reconstruction of graph signals can typically be categorized to two main groups of selection sampling \cite{chen2015discrete} and aggregation sampling \cite{marques2016sampling}, where the focus of this paper is on the former. 

Sampling of noise-corrupted signals using randomized schemes including uniform and leverage score sampling is studied in \cite{chen2015discrete,chen2016signal}, for which optimal selection sampling distributions and 
performance bounds are derived. Reference \cite{puy2016random} borrows the variable density sampling strategy from compressed sensing to derive a random selection sampling scheme with optimal distribution.
State-of-the-art random selection sampling schemes typically require sampling more than $k$ nodes of a $k$-band-limited graph signal to achieve near perfect recovery and their performance deteriorates for reconstruction of graph signals with relatively large bandwidth. 

A main challenge in sampling and reconstruction is the problem of identifying the support of bandlimited graph signals  \cite{marques2016sampling,di2016adaptive,romero2017kernel,narang2013signal,anis2016efficient}.
In \cite{narang2013signal,anis2016efficient}, support identification of smooth graph signals is studied. However, Approaches in \cite{narang2013signal,anis2016efficient} rely solely on a user-defined sampling strategy and the graph Laplacian, and disregard the availability of observations of the graph signal. A similar scheme is developed in \cite{marques2016sampling} for aggregation sampling where under established assumptions on topology of the graph, conditions for exact support identification from noiseless measurements are put forth. An alternating minimization approach is proposed in \cite{di2016adaptive} that jointly recovers the unknown support of the signal and designs a sampling strategy in an iterative fashion. However, the convergence of the alternating scheme in \cite{di2016adaptive} is not guaranteed and the conditions for exact support identifications are unknown \cite{di2016adaptive}.

In this work, we consider the task of selection sampling and reconstruction of bandlimited graph signals with unknown support.
Following ideas from compressed sensing, we  propose a novel and efficient iterative selection sampling approach that exploits the low-cost selection criterion of the orthogonal matching pursuit algorithm \cite{pati1993orthogonal} to recursively select a subset of nodes of the graph. We theoretically demonstrate that in the noiseless case the original $k$-bandlimited signal is exactly recovered from the set of selected nodes with cardinality $k$. When the measurements are noisy, we establish a worst-case performance bound on the reconstruction error of the proposed algorithm. We further extend our results to the case of bandlimited signals with unknown supports, and demonstrate that under a pragmatic SNR condition, the proposed framework still requires $k$ samples to ensure exact recovery of signals with unknown supports from historical samples of the graph signal.
Simulation studies show the proposed sampling algorithm compares favorably to competing random selection sampling alternatives.\footnote{Proofs of the theoretical results in this paper are omitted for brevity and stated in the extended manuscript \cite{exman}.}

\vspace{-0.2cm}
\section{Preliminaries}\label{sec:pre}
Consider a network represented by a graph $\mathcal{G}$ consisting of a node set $\mathcal{N}$ of 
cardinality $N$ and a weighted adjacency matrix $\mathbf{A} \in \mathbbm{R}^{N \times N}$ whose
$(i,j)$ entry, $\A_{ij}$, denotes weight of the edge connecting node $i$ to node $j$.  A \textit{graph 
	signal} $\mathbf{x} \in \mathbbm{R}^{N}$ is a vertex-valued network 
process that can be represented by a vector of size $N$ supported on $\mathcal{N}$, where its $i\ts{th}$ component denotes the signal value at node $i$.
Let $\x$ be a graph signal which is $k$-bandlimited in a given basis $\mathbf{V} \in \mathbbm{R}^{N \times N}$. This means that the signal's so-called graph Fourier transform (GFT) $\bar{\x} = \V^{-1} \x$ is $k$-sparse. There are several choices for $\mathbf{V}$ in the literature with most aiming to decompose a graph signal into different modes of variation with respect to the graph topology. For instance, $\mathbf{V} = [\mathbf{v}_{1},\cdots,\mathbf{v}_{N}]$ can be defined via the Jordan decomposition of the adjacency matrix \cite{DSP_freq_analysis,deri2017spectral}, through the eigenvectors of the Laplacian when $\mathcal{G}$ is undirected \cite{shuman2013}, or it can be obtained as the result of an optimization procedure \cite{shafipour2017digraph,sardellitti}. In this paper, we assume the adjacency matrix $\A = \V \mathbf{\Lambda}\V^{-1}$ is normal which in turn implies $\V$ is unitary and $\V^{-1} = \V^\top$.
%
%

Recall that since $\x$ is bandlimited, $\bar{\x}$ is sparse with at most $k$ nonzero entries. Let $\K$ be the support set of $\bar{\x}$, where $|\K| = k$. Then, one can write $\x = \U\bar{\x}_\K$, where $\U = \V_{\K,c}$ and $\V_{\K,c}$ ($\x_\K$) is a submatrix (subvector) of $\V$ ($\x$) that
contains columns (elements) indexed by the set $\K$. Also, notation $\V_{\K,r}$ will represent the submatrix that contains rows indexed by $\K$. In the sequel, we first assume that the support set $\K$ is known. In section \oldref{sec:rec}, we discuss how to tackle sampling scenarios where $\K$ is unknown.

\vspace{-0.2cm}
\section{Proposed Framework}\label{sec:bl}
In this section, we consider sampling of bandlimited signals with known support. Specifically, we assume that a graph signal $\x$ is sparse given a basis $\V$. Let $\A = \V \mathbf{\Lambda}\V^\top$ be the normal decomposition of $\A$, the adjacency matrix of the undirected graph $\mathcal{G}$. we first consider the noise-free scenario and then extend our results to the case of sampling and reconstruction from noisy measurements.
\subsection{Sampling bandlimited graph signals}
In selection sampling (see, e.g.\cite{chen2015discrete}),  sampling a
graph signal amounts to finding a sampling matrix $\C \in \{0,1\}^{k\times N}$, such that $\tilde{\x} = \C\x$, where $\tilde{\x}$ is the sampled graph signal. Since $\x$ is bandlimited with support $\mathcal{K}$, and $\x = \U\bar{\x}_\K$, it holds that $\tilde{\x} = \C\U\bar{\x}_\K$.
The original signal can then be reconstructed according to 
\begin{equation}\label{eq:rec1}
\hat{\x} =\U\bar{\x}_\K= \U(\C\U)^{-1}\tilde{\x}.
\end{equation}
According to \eqref{eq:rec1}, a necessary and sufficient condition for perfect reconstruction, i.e. $\hat{\x} = \x$ is invertibility of matrix $\C\U$. However, as argued in \cite{marques2016sampling,shomorony2014sampling} (see, e.g., Section III-A in \cite{marques2016sampling}), current selection sampling approaches cannot construct a sampling matrix to ensure $\C\U$ is invertible for an arbitrary graph, and invertibility of  $\C\U$ is checked by inspection which requires intensive computational costs for large graphs.
To overcome this issue, motivated by the well-known OMP algorithm in compressed sensing \cite{pati1993orthogonal}, we propose a simple iterative scheme with complexity $\mathcal{O}(Nk^2)$ that guarantees perfect recovery of $\x$ from the sampled signal $\tilde{\x}$.

The proposed approach (see Algorithm \ref{alg:1}) works as follows. First, the algorithm chooses an arbitrary node of the graph with index $\ell$ for some $\ell \in [N]$ as a {\it residual node}. Then, in $i\ts{th}$ iteration, the algorithm identifies a node with index $j_s$ to be included in the sampling set $\S$ according to 
\begin{equation}
	j_s = \argmax_{j \in \mathcal{N}\backslash \S}\frac{|\r_{i-1}^\top \u_j|^2}{\|\u_j\|_2^2},
\end{equation}
where $\r_i = \P_\S^\bot\u_\ell$ is a residual vector initialized at $\r_0 = \u_\ell$, $\P_\S^\bot\I_N-\U_{\S,r}^\top (\U_{\S,r}^\top)^\dagger$ is the projection operator onto the orthogonal complement of the subspace spanned by
the rows of $\mathbf{U}_{\S,r}$, and $\U_{\S,r}^\dagger=\left(\U_{\S,r}^{\top}\U_{\S,r}\right)^{-1}\U_{\S,r}^{\top}$ denotes the Moore-Penrose pseudo-inverse of $\U_{\S,r}$ . This procedure is repeated for $k$ iterations to construct $\S$.
\renewcommand\algorithmicdo{}	
\begin{algorithm}[t]
	\caption{Iterative Selection Sampling}
	\label{alg:1}
	\begin{algorithmic}[1]
		\STATE \textbf{Input:} $\U$, $k$, number of samples $m\geq k$.
		\STATE \textbf{Output:} Subset $S\subseteq \mathcal{N} $ with $|\S|=m$.
		\STATE Initialize $\S =  \emptyset$, $\r_0 = \u_\ell$ for some $\ell \in [N]$, $i = 0$.
		\WHILE{$|\S|<m$}\vspace{0.1cm}
		\STATE $i\leftarrow i+1$
		\STATE $i_s = \argmax_{j \in \mathcal{N}\backslash \S}\frac{|\r_{i-1}^\top \u_j|^2}{\|\u_j\|_2^2}$\vspace{0.1cm}
		\STATE Set $\S \leftarrow \S\cup \{i_s\}$\vspace{0.1cm}
		\STATE $\r_i = \P_\S^\bot\u_\ell$
		\ENDWHILE
		\RETURN $\S$.
	\end{algorithmic}
\end{algorithm}
Theorem \ref{thm:iss} demonstrates that Algorithm \ref{alg:1} returns a sampling set that ensure perfect recovery of the graph signal $\x$ under noise-free scenario.
\begin{theorem}\label{thm:iss}
Let $\S$ be the sampling set constructed by Algorithm \ref{alg:1} and let $\C$ be the corresponding sampling matrix such that $|\S| = k$. Then, the matrix $\C\U$ is invertible (with probability one).
\end{theorem}
Theorem \ref{thm:iss} states that as long as the adjacency matrix $\A$ is normal, the proposed selection sampling scheme guarantees perfect reconstruction of the original signal from its noiseless samples. This condition guarantees recovery for a wide class of graph structures in compared to, for example, the aggregation sampling scheme \cite{marques2016sampling} that requires eigenvalues of $\A$ that are indexed by $\K$ to be distinct.
\subsection{Sampling in the presence of noise}
We now provide an extended analysis of the proposed Algorithm~\ref{alg:1} for scenarios where only noisy samples of the graph signal are available. Note that because of the noise, a perfect reconstruction is not possible. Nonetheless, we provide an upperbound on the reconstruction error of the proposed sampling scheme as a function of the noise covariance and the sampling matrix $\C$. Another different aspect of sampling and reconstruction under noise is that it might be desirable to select $m\geq k$ nodes as the sampling set to achieve better reconstruction accuracy. This is in contrast to the noiseless case where only $m= k$ is sufficient for perfect reconstruction if the sampling set is constructed by Algorithm \ref{alg:1} as stated in Theorem~\ref{thm:iss}.

To further understand, let $\y = \x + \n$ be the noise-corrupted signal, where $\n \in \R^{N}$ is the zero-mean noise vector with covariance matrix $\E[\n\n^\top]=\Q$.
Therefore, since $\x = \U\bar{\x}_\K$, the samples $\tilde{\x}$ and the non-zero frequency components of $\x$ are related via the linear model
\begin{equation}\label{eq:modeln}
\tilde{\x} = \y_\S = \U_{\S,r}\bar{\x}_\K +\n_{\S},
\end{equation}
where $\U_{\S,r} = \C\U$, $\y_\S = \C\y$, and  $\n_\S = \C\n$. The reconstructed signal in the Fourier domain satisfies the normal equation \cite{kay1993fundamentals},
\begin{equation}\label{eq:normal}
\U_{\S,r}^\top\Q_\S^{-1}\U_{\S,r}\hat{\bar{\x}} = \U_{\S,r}^\top \Q_\S^{-1} \tilde{\x},
\end{equation}
where $\Q_\S = \C\Q\C^\top$ is the covariance of $\n_\S$. A necessary condition for recovery is invertibility of the matrix $\U_{\S,r}^\top\Q_\S^{-1}\U_{\S,r}$. Indeed, as stated in the following theorem, if $\S$ is selected using Algorithm \ref{alg:1},  $\U_{\S,r}^\top\Q_\S^{-1}\U_{\S,r}$ is invertible and we can recover the original graph signal, up to an error term.
\begin{proposition}\label{thm:nois}
Let $\S$ be the sampling set constructed by Algorithm \ref{alg:1}, $\C$ be the corresponding sampling matrix and denote $\U_{\S,r} = \C\U$. Then, matrix $\U_{\S,r}^\top\Q_\S^{-1}\U_{\S,r}$ is invertible and if $\|\n\|_2 \leq \epsilon_\n$, then reconstruction error of the signal reconstructed from $\S$ satisfies
\begin{equation}\label{eq:boundn}
\|\hat{\x}-\x\|_2 \leq \sigma_{\max}((\U_{\S,r}^\top\Q_\S^{-1}\U_{\S,r})^{-1}\U_{\S,r}^\top \Q_\S^{-1})\epsilon_\n,
\end{equation}
where $\sigma_{\max}(.)$ outputs the maximum singular value of a matrix.
\end{proposition}

Proposition \ref{thm:nois} states an explicit bound on the reconstruction error of the proposed sampling scheme for general noise models with bounded $\ell_2$-norm. Also, the proposed selection sampling scheme preserves the structural properties of noise's statistics. More specifically, if $\n$ is white with $\Q = \sigma^2\I_N$, the effective noise remains unchanged and hence white which is in contrast to, for example, aggregation sampling that the effective noise becomes correlated; see Section IV-A in \cite{marques2016sampling}.
 \begin{figure*}[t]
	\begin{minipage}[t]{0.41\textwidth}
		\centering
		\includegraphics[width=1\linewidth]{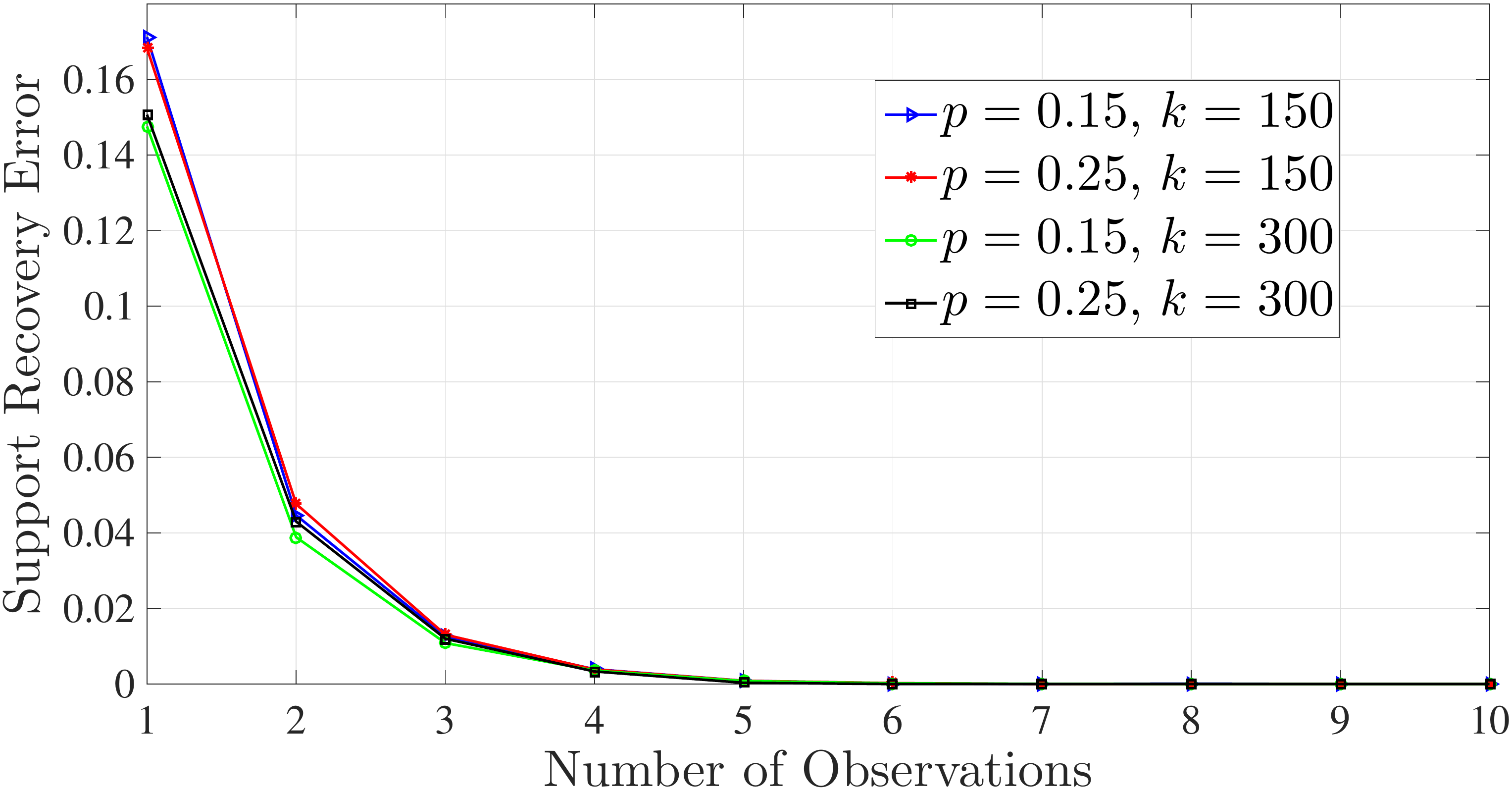}
		\centerline{\footnotesize(a)}\medskip
	\end{minipage}
	\hfill
	\begin{minipage}[t]{0.29\textwidth} 
		\centering
		\includegraphics[width=1\linewidth]{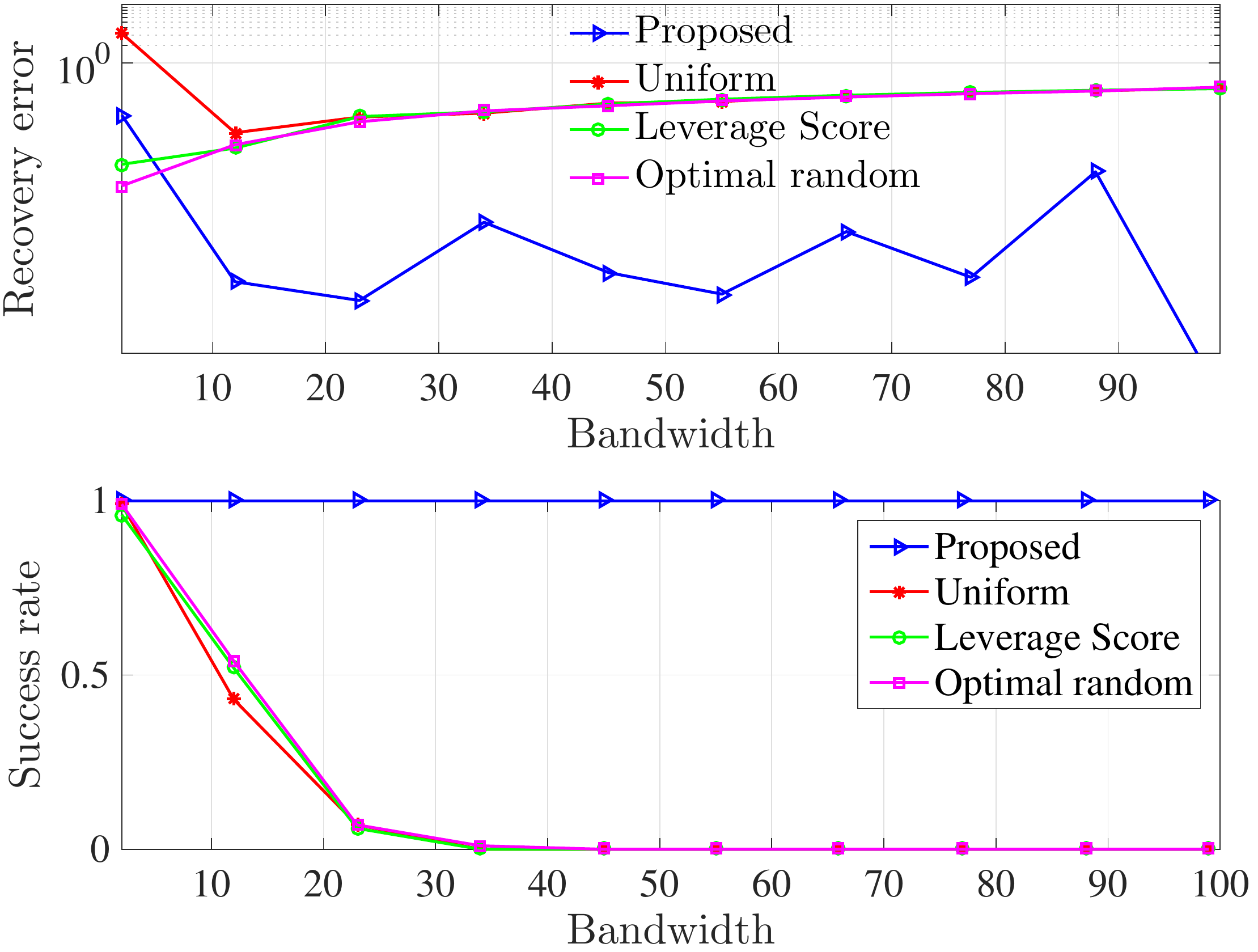}
		\centerline{\footnotesize(b)}\medskip
	\end{minipage}
	\hfill
	\begin{minipage}[t]{0.29\textwidth} 
		\centering
		\includegraphics[width=1\linewidth]{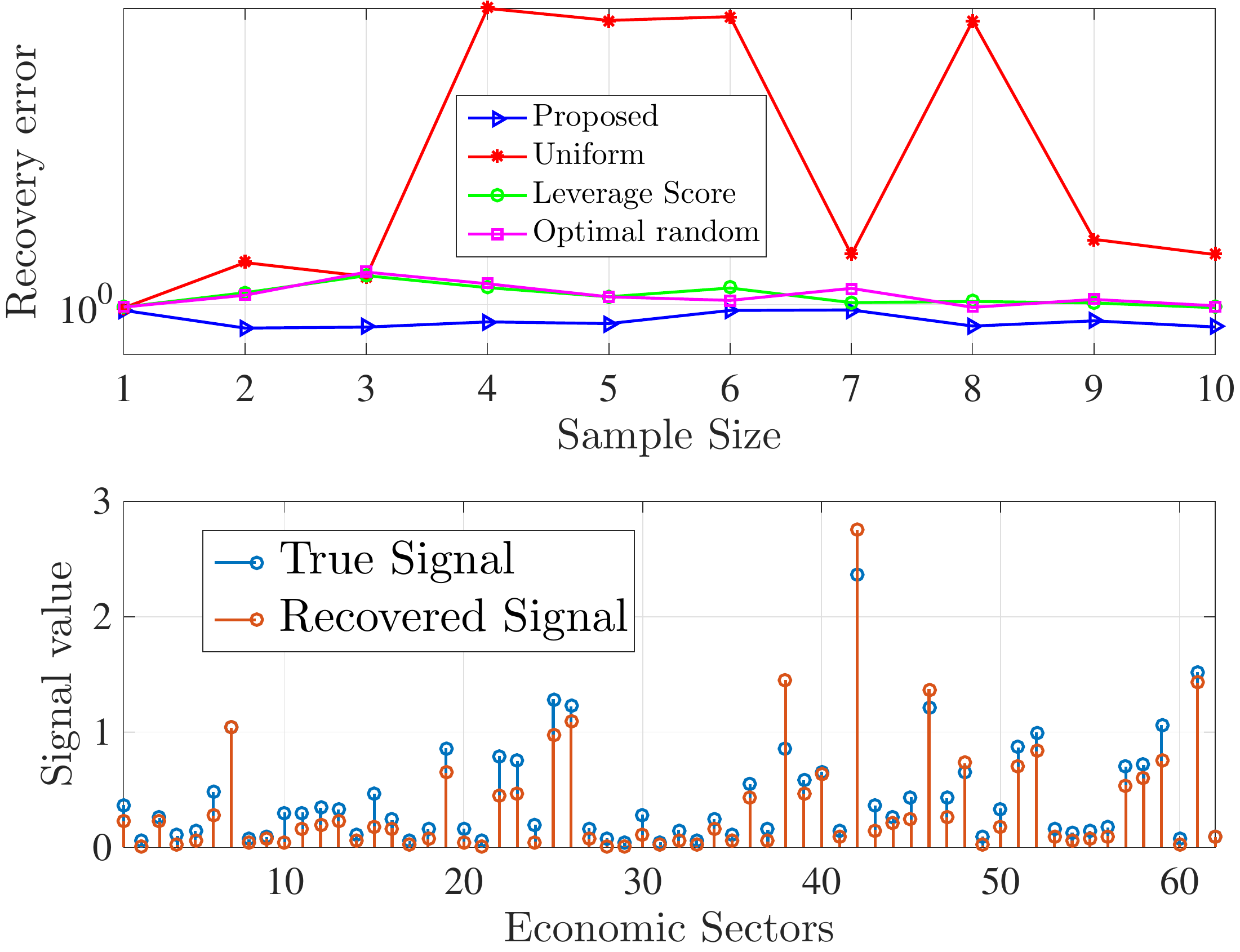}
		\centerline{\footnotesize (c)}\medskip
	\end{minipage}
	\vspace{-0.4cm}
	\caption{(a) Support recovery error of the proposed scheme versus number of observations for different degrees of connectivities ($p$) and bandwidths ($k$). (b) Recovery error (top) and success rate (bottom)  comparison of Algorithm 1 and different random selection sampling schemes versus bandwidth ($k$) for undirected Erd\H{o}s-R\'enyi random graphs. (c) Top: Recovery error comparison of different selection sampling schemes as a function of sample size for the economy network. Bottom: Recovered and true graph signal at different economic sectors using Algorithm 1.}
	\vspace{-0.3cm}
\end{figure*}
\vspace{-0.2cm}
\section{Support recovery from historical signals}\label{sec:rec}
So far, we have assumed that the support of the bandlimited graph signal $\x$ is known. However, in many practical applications the support of the signal might be unknown and one needs to recover the support prior to or concurrent with the  sampling step. In this section, we address the problem of identifying the support of bandlimited graph signals from a minimal number of (fully) observed signals at all nodes prior to sampling.
Note that cost of evaluating the signal value at all nodes of the graph stems from accommodating identification of unknown supports without making any specific assumptions on the structure of the graph. 

To that end, for a given graph and its Fourier basis $\mathbf{V}$, suppose that we observe $P$ signals collected in a matrix $\mathbf{Y} = [\mathbf{y}^{1}, \cdots, \mathbf{y}^{P}] \in \mathbb{R}^{N \times P}$. The goal is to perform support recovery of the underlying bandlimited signals collected in $\mathbf{X} = [\mathbf{x}^{1}, \cdots, \mathbf{x}^{P}] \in \mathbb{R}^{N \times P}$ which amounts to estimating sparse GFTs $\bar{\mathbf{X}} = [\bar{\x}^{1},\cdots,\bar{\x}^{P}] = \mathbf{V}^{T} \mathbf{X}$. Pragmatically, we assume that $\mathbf{V}$ has full rank and the GFTs $\bar{\mathbf{x}}^{i}$ share a common support. The latter assumption specifies matrix $\bar{\mathbf{X}}$ in a block-sparse model which has entire rows as zeros or nonzeros. Recall that $\mathcal{K}$ is the support set of GFTs $\bar{\mathbf{x}}^{i}$ with $|\mathcal{K}| = k$. Then, upon defining
\begin{equation} \label{e:def_block_sparse}
\begin{aligned}
\mathit{\Psi_{\mathcal{K}}} := & \big\{ \bar{\mathbf{X}} = [\bar{\x}^{1},\cdots,\bar{\x}^{P}] \in \mathbb{R}^{N \times P} \quad \text{such that} \\ & \bar{\mathbf{X}}_{\mathcal{K},r} = \mathbf{0} \hspace{0.1cm} \text{for}  \hspace{0.1cm} i \notin \mathcal{K}, \mathcal{K} \subset [N], \lvert \mathcal{K} \rvert = k  \big\},
\end{aligned}
\end{equation}
the bandlimited signal recovery (thus support $\mathcal{K}$) boils down to solving
\begin{equation}
\begin{aligned}
\label{e:opt_support}
& \underset{\mathbf{\bar{X}}}{\text{min}}
& &  \lVert \mathbf{\bar{X}} - \mathbf{\bar{Y}} \rVert_{F}^{2} 
& \text{s.t.}
& &  \mathbf{\bar{X}}\in \mathit{\Psi_{\mathcal{K}}},
\end{aligned}
\end{equation}
where $\bar{\Y} = \V^{\top}\Y $. Note that support recovery is a byproduct of finding the optimum in \eqref{e:opt_support}. Inspired by \cite{baraniuk2010model}, we propose to use $(2,0)$ mixed norm to reformulate \eqref{e:opt_support} as
\begin{equation}
\begin{aligned}
\label{e:opt_support_mixed_norm}
& \underset{\bar{\mathbf{X}}}{\text{min}}
& &  \lVert \bar{\mathbf{X}} - \bar{\mathbf{Y}} \rVert_{F}^{2} 
& \text{s.t.}
& &  \lVert \bar{\mathbf{X}} \rVert_{2,0} \leq k,
\end{aligned}
\end{equation}
where the $(p,q)$ mixed norm of matrix $\bar{\mathbf{X}}$ is defined as
\begin{equation} \label{e:mixed_norm}
\lVert \bar{\mathbf{X}} \rVert_{p,q} := \big(\sum_{i=1}^{N} \lVert \bar{\x}^{i} \rVert_{p}^{q} \big)^{1/q}.
\end{equation}
Note that $(2,0)$ norm counts the number of nonzero rows in $\bar{\mathbf{X}}$.

It is immediately apparent that the solution $\bar{\mathbf{X}}^{\star}$ of \eqref{e:opt_support_mixed_norm} is obtained by the row-wise $l_{2}$ norm thresholding on $\bar{\mathbf{Y}}$. Specifically, upon calculating the $l_{2}$ norm of the rows of $\bar{\mathbf{Y}}$ (i.e., $\lVert \bar{\mathbf{y}}(i,:) \rVert_{2}$ for all $i$, where $\bar{\mathbf{Y}}(i,:)$ denotes the $i$th row of $\bar{\mathbf{Y}}$), let $\zeta$ be the $k^{th}$ largest $l_{2}$ norm. Then, the solution $\bar{\mathbf{X}}^{\star}$ has rows given by

\begin{align} \label{e:support_recovery_solution_form}
\bar{\mathbf{X}}(i,:)^{\star}=
\begin{cases}
\bar{\mathbf{Y}}(i,:) \quad \lVert \bar{\mathbf{y}}(i,:) \rVert_2 \geq \zeta\\
\mathbf{0} \quad \quad \quad \quad \quad \text{o.w.}
\end{cases}
\end{align}
The running time of finding the solution of \eqref{e:opt_support_mixed_norm} according to \eqref{e:support_recovery_solution_form} is $\mathcal{O}(NP+N \text{log} N)$. This comprises of finding the $l_{2}$ norm of the rows in $\mathcal{O}(NP)$ and then performing an off-the-shelf sorting algorithm (e.g., merge sort) in $\mathcal{O}(N \text{log} N)$.

Analyzing support identifiability using the proposed scheme is challenging in the presence of noise. To that end, we impose the energy constraint on the noise signals so not to exceed a threshold $\epsilon_{\n}$. Specifically, in the following theorem, we provide sufficient conditions to exactly recover the support from the noisy observations under the energy-constrained noise model.
\begin{theorem}\label{thm:supn}
	Consider support recovery of $P$ bandlimited graph signals $\mathbf{X} = \mathbf{V} \mathbf{\bar{X}} = \mathbf{V} [\bar{\x}^{1}, \cdots, \bar{\x}^{P}] \in \mathbb{R}^{N \times P}$ with unknown shared support $\mathcal{K}$ with $\lvert \mathcal{K} \rvert = k$; i.e., $\mathbf{\bar{X}} \in \mathit{\Psi_{\mathcal{K}}}$. Let $\mathbf{{Y}} = [\y^{1}, \cdots, \y^{P}] = \mathbf{X} + \mathbf{N}$ denote the noisy measurements of graph signals $\mathbf{X}$, where $\mathbf{{N}} = [\n^{1}, \cdots, \n^{P}] \in \mathbb{R}^{N \times P}$ models the corruption noise signals. Assume $\mathbf{V}$ is orthogonal and $\n^{i}$s have bounded $l_{2}$ norm; in particular $\lVert \n^{i} \rVert_2 \leq \epsilon_{\n}$ for all $i \in [P]$. Then, GFTs $\mathbf{\bar{X}}$ and, as a byproduct, the support $\mathcal{K}$ are identifiable in \eqref{e:opt_support} if
	\begin{equation} \label{e:support_recovery_condition}
	\underset{i \in \mathcal{K}}{\text{min}} \quad \lVert \bar{\mathbf{x}}(i,:) \rVert_2 > 2 \epsilon_{\n} \sqrt{P}.
	\end{equation}
\end{theorem}

Inspection of \eqref{e:support_recovery_condition} shows that in the noiseless scenario (i.e., $\epsilon_{\n} = 0$), the support is identifiable even for $P=1$. Moreover, in the presence of noise, it is conceivable that as $P$ grows, the chance of \eqref{e:support_recovery_condition} being satisfied increases and the support recovery problem is rendered identifiable for more observations; see also the numerical tests in Section~\ref{sec:sim}.
Naturally, another insightful identifiability result for stochastic noise models would be valuable, but left as future work.

Compared to existing sampling schemes capable of support identification in noiseless scenario, e.g. \cite{marques2016sampling} that require twice as many samples as the bandwidth of the graph signal (i.e., $k$) for perfect reconstruction of $\x$ and further conditions on the structure of the graph -- e.g., distinct nonzero eigenvalues of the graph shift operator (adjacency $\A $ or  the Laplacian $\mathbf{L}$ defined as $\mathbf{L} = \text{diag}(\mathbf{A}\mathbf{1}_{N}) - \mathbf{A}$, for example) -- the proposed framework [cf. \eqref{e:support_recovery_solution_form}] needs only $k$ samples of the nodes of the graph to achieve perfect recovery and identification of $\K$. Additionally, the proposed framework is capable of support identification for the broad range of graphs which have normal graph shift operator. Adaptive sampling and reconstruction of graph signals with unknown supports can be achieved as a result of  combining  the proposed support identification framework with the proposed sampling scheme outlined in Algorithm 1, and is left as a subject of our future studies.

\vspace{-0.2cm}
\section{Simulation Results}\label{sec:sim}
We study the recovery of signals supported on synthetic and real-world graphs to assess performance of the proposed support recovery and sampling algorithms. To this end, we first consider undirected Erd\H{o}s-R\'enyi random graphs $\mathcal{G}$ of size $N = 1000$ and edge probability $0.15$ or $0.25$ \cite{newman2010networks}. Bandlimited graph signals $\mathbf{x} = \mathbf{U}\bar{\mathbf{x}}_{K}$ are generated by taking $\mathbf{U}$ as the $k$ randomly selected eigenvectors $\mathbf{V}$ of the graph adjacency matrix, where $k=150$ or $300$. The non-zero frequency components $\bar{\mathbf{x}}_{K}$ are drawn independently from a zero-mean Gaussian distribution with standard deviation $100$. We also corrupt the signals (measurements) by additive Gaussian noise with $20$dB power. We first start by observing $P$ signals across all nodes when the frequency support is unknown and try to recover the support using the proposed formulation in \eqref{e:opt_support_mixed_norm}, \eqref{e:support_recovery_solution_form}. Fig. 1(a) depicts the support recovery error as the ratio of number of elements common in the ground truth and inferred frequency support to the bandwidth ($k$) as a function of $P$, where the results are obtained by averaging over $100$ Monte-Carlo simulations. We notice that as $P$ increases, the recovery error decreases monotonically [cf. Theorem \ref{thm:supn}] for different degrees of connectivity ($p$) and bandwidths ($k$). As expected, the recovery performance does not rely on the edge probabilities, since we only need $\mathbf{V}$ to be full rank which is satisfied in all undirected graphs considered in this simulation study. Moreover, since with higher bandwidth the energy in GFT components and the chance of satisfying \eqref{e:support_recovery_condition} increases, the support recovery task becomes easier, as predicted by the results of Theorem \ref{thm:supn}.

Next, we consider the task of sampling and reconstruction of noise-corrupted bandlimited graph signals with known support. Specifically, we consider undirected Erd\H{o}s-R\'enyi random graphs $\mathcal{G}$ of size $N = 100$ and edge probability $0.2$. We generate $\mathbf{x} = \mathbf{U}\bar{\mathbf{x}}_{K}$ by taking $\mathbf{U}$ the first $k$ eigenvectors $\mathbf{V}$ of the graph adjacency matrix, where we vary $k$ linearly from $2$ to $99$. The non-zero frequency components $\bar{\mathbf{x}}_{K}$ are drawn independently from a zero-mean Gaussian distribution with standard deviation $100$ and the signal is corrupted with a Gaussian noise term with $\mathbf{Q} = 0.02^2\mathbf{I}_N$. We compare the recovery performance of the proposed scheme in Algorithm~\ref{alg:1} with the state-of-the-art uniform, leverage score, and optimal random sampling schemes \cite{chen2015discrete,chen2016signal,puy2016random}. We define the recovery error as the ratio of the error energy to the true signal's energy.  Furthermore, {\it success rate} \cite{chen2015discrete} is defined as the fraction of instances where $\C\U$ is invertible [cf. \eqref{eq:rec1}]. The results, averaged over 100 independent instances, are illustrated in Fig 1(b). As we can see from Fig 1(b)(top), the proposed scheme consistently achieves a lower recovery error compared to competing schemes. Moreover, as Fig 1(b)(bottom) illustrates, when the bandwidth increases success rate of random sampling schemes decreases while the success rate of proposed scheme is always one, as we proved in Theorem \ref{thm:iss}. 

Finally, we use the data from Bureau of Economic Analysis of the U.S. Department of Commerce which publicizes an annual table of input and outputs organized by economic sectors \footnote{Dataset from https://www.bea.gov}. Specifically, we use $62$ industrial sectors as defined by the North American Industry Classification System as nodes and construct the weighted edges and the graph signal similar to \cite{marques2016sampling}. To that end, the (undirected) edge weight between the two nodes represents the average total production of the sectors, the first sector being used as the input of the other sector, expressed in trillions of dollars per year. This edge weight is averaged over the years $2008$, $2009$, and $2010$. Also, two artificial nodes are connected to all $62$ nodes as the added value generated and the level of production destined to the market of final users. Thus, the final graph has $N = 64$ nodes. The weights lower than $0.01$ are then thresholded to zero and the eigenvalue decomposition of the corresponding adjacency matrix $\mathbf{A} = \mathbf{V} \mathbf{\Lambda} \mathbf{V}^{\top}$ is obtained. A graph signal $\mathbf{x} \in \mathbb{R}^{64}$ can be regarded as a unidimensional total production -- in trillion of dollars -- of each sector during the year 2011. Signal $\mathbf{x}$ is shown to be approximately (low-pass) bandlimited in \cite[Fig. 4(a)(top)]{marques2016sampling} with a bandwidth of $4$.

We try to interpolate sectors' production by observing a few nodes using Algorithm~\ref{alg:1} assuming that the signal is low-pass (i.e., with smooth variations over the built network). Then, we vary the sample size and compare the recovery performance of the proposed scheme with the state-of-the-art uniform, leverage score, and optimal random sampling schemes \cite{chen2015discrete,chen2016signal,puy2016random} averaged over $1000$ Monte-Carlo simulations as shown in Fig. 1(c)(top). As the figure indicates, the proposed algorithm outperforms uniform, leverage score, and optimal random sampling schemes\cite{chen2015discrete,chen2016signal,puy2016random}. However, we are not experiencing perfect recovery for our proposed Algorithm~\ref{alg:1} in this noiseless scenario because the signal is not purely bandlimited.
Moreover, Fig. 1(c)(bottom) shows a realization of the graph signal $\mathbf{x}$ superimposed with the reconstructed signal obtained using Algorithm~\ref{alg:1} with $k=2$ for all nodes excluding two artificial ones. 
The recovery error of the reconstructed signal is approximately $1.32\%$ and as Fig. 1(c)(bottom) illustrates,  $\hat{\mathbf{x}}$ closely approximates $\mathbf{x}$.
\vspace{-0.2cm}
\section{Acknowledgment} We would like to thank the authors in \cite{marques2016sampling} for providing the data used for the economy network in the last experiment.
\vspace{-0.2cm}
\section{Conclusion} \label{sec:concl}
We considered the task of sampling and reconstruction of $k$-bandlimited graph signals.
We  proposed an efficient iterative sampling approach that exploits the low-cost selection criterion of the orthogonal matching pursuit algorithm to recursively select a subset of nodes of the graph. We also theoretically showed that in the noiseless case the original $k$-bandlimited signal is exactly recovered from the set of selected nodes with cardinality $k$. In the case of noisy measurements, we established a worst-case performance bound on the reconstruction error of the proposed algorithm. We further extended our results to the case where the support of the bandlimited signal is unknown and demonstrated under a mild SNR condition, the proposed framework still requires $k$ samples to ensure exact recovery of signals with unknown supports from historical samples of the graph signal.
Simulation studies showed that the proposed sampling algorithm compares favorably to competing alternatives.

\clearpage
\bibliographystyle{ieeetr}\footnotesize
\bibliography{refs}
\end{document}